\documentclass[12pt]{iopart}

\usepackage{graphicx}
\usepackage{subfigure}
\usepackage{bm}
\usepackage{hyperref}

\def\simleq{\; \raise0.3ex\hbox{$<$\kern-0.75em \raise-1.1ex\hbox{$\sim$}}\; }
\def\simgeq{\; \raise0.3ex\hbox{$>$\kern-0.75em \raise-1.1ex\hbox{$\sim$}}\; }

\newcommand{\MV}{{\rm MV}}
\newcommand{\GeV}{{\rm GeV}}
\newcommand{\TeV}{{\rm TeV}}

\newcommand{\kpc}{{\rm kpc}}
\newcommand{\pc}{{\rm pc}}
\newcommand{\m}{{\rm m}}
\newcommand{\cm}{{\rm cm}}

\newcommand{\s}{{\rm s}}

\newcommand{\sr}{{\rm sr}}

\begin{document}
\title[Cosmic Ray Nuclei in the Galaxy] {Cosmic-Ray Nuclei, Antiprotons and Gamma-rays in the Galaxy: a New Diffusion Model}

\author{Carmelo Evoli$^1$, Daniele Gaggero$^{2,3}$, Dario Grasso$^3$, Luca Maccione$^{1,4}$}
\address{$^1$ SISSA, via Beirut, 2-4, I-34014 Trieste}
\address{$^2$ Dipartimento di Fisica ``E. Fermi", Universit\`a di Pisa, Largo B. Pontecorvo, 3, I-56127 Pisa}
\address{$^3$ INFN, Sezione di Pisa, Largo B. Pontecorvo, 3, I-56127 Pisa}
\address{$^4$ INFN, Sezione di Trieste, Via Valerio, 2, I-34127 Trieste}

\eads{\mailto{evoli@sissa.it}, \mailto{dario.grasso@pi.infn.it}, \mailto{maccione@sissa.it},
\mailto{daniele.gaggero@pi.infn.it}}

\begin{abstract}
We model the transport of cosmic ray nuclei in the Galaxy by means of a new numerical code. Differently from previous numerical models we account for a generic spatial distribution of the diffusion coefficient. We found that in the case of radially uniform diffusion, the main secondary/primary ratios (B/C, N/O and sub-Fe/Fe) and the  modulated antiproton spectrum match consistently the available observations. Convection and re-acceleration do not seem to be required in the energy range we consider: $1 \le E \le 10^3~\GeV/{\rm nucleon}$. We generalize these results accounting for radial dependence of the diffusion coefficient, which is assumed to trace that of the cosmic ray sources. While this does not affect the prediction of secondary/primary ratios, the simulated longitude profile of the diffuse $\gamma$-ray emission is significantly different from the uniform case and may agree with EGRET measurements without invoking {\it ad hoc} assumptions on the galactic gas density distribution.

\end{abstract}

\section{Introduction}

Understanding the origin and propagation of Cosmic Rays (CR) in the Galaxy is an intriguing puzzle which requires the combination of many different observations over a wide range of energies. While simplified propagation models, most noticeably the {\it leaky-box} and the modified slab models, succeeded interpreting the main features of the CR nuclear composition and energy spectra for $1 \simleq E \simleq 100~\GeV/{\rm n}$\footnote{This unit of measure corresponds to the energy $E$ competing on average to each nucleon within a nucleus with $A$ nucleons and energy $A\times E$.}, more realistic diffusion models are called for to provide a comprehensive
description of multi-channel observations (including heavy nuclei, electrons,  $\gamma$-rays and antimatter particles) accounting for the growing amount of available astrophysical data  (see \cite{Strong:2007nh}  for a comprehensive review).

Two main approaches have been developed so far: analytical (or semi-analytical) diffusion models (see e.g.~\cite{Berezinsky:book} and ref.s therein), which solve the CR transport equation by assuming simplified distributions for the sources and the  interstellar gas, and fully numerical diffusion models. Well known recent realizations of those two approaches are respectively the {\it two-zone model} \cite{Maurin:01,Maurin:02} and the GALPROP code \cite{Strong:98,Strong:04,GALPROPweb}. In the case of GALPROP, the adoption of  realistic gas and radiation field distributions allows to model also the spectrum and angular distribution of the $\gamma$-ray secondary emission.

Although these models represent a significant step forward with respect to previous analyses, they still perform a number of simplifications with respect to a more realistic physical scenario. Most noticeably, they assume diffusion to be statistically isotropic and homogenous, i.e.~they adopt the same diffusion coefficient value (or at most two values in the two-zone model) all over the propagation volume.

However, such assumptions may not always be justified, as diffusion coefficients generally depend on the regular magnetic field orientation and on the ratio between the regular and chaotic magnetic field energy densities. Although these quantities are poorly known, several observations and theoretical arguments suggest that they are far from being spatially homogeneous in the Galaxy (see sec.~\ref{sec:diffusion} for more details). This may have relevant consequences for the CR spatial distribution in the Galaxy, for the angular distribution of the secondary $\gamma$-ray and neutrino emissions \cite{Evoli:2007iy} and to interpret the CR anisotropy.

In order to be able to test some of those effects, as well as to verify previous results which have been derived in the literature under more conventional conditions, 
%
%
we developed a new numerical code, DRAGON (Diffusion of cosmic RAys in Galaxy modelizatiON). 
The main focus of the project is on CR transport. The code solves a general version of the  diffusion equation allowing for position-dependent diffusion; the solver is linked to routines and data tables taken from the current public version of GALPROP~\footnote{Galprop code can be downloaded from:~\url{http://galprop.stanford.edu}}. (see the text for more details)

In its present version it allow to model CR nuclei transport at energies $E_{\rm min}\simgeq 1~\GeV$/n as well as the secondary $\gamma$-ray and neutrino emission produced by their interaction with the interstellar medium (ISM). In this work we disregard CR convection and re-acceleration (i.e.~we work in a plain diffusion (PD) regime) and show that most relevant measurements can be reproduced under these conditions. Above $E_{\rm min}$ we expect that no other physical input than source spectra, diffusion and fragmentation processes can determine secondary/primary ratios, hence a comparison of our prediction with experimental data should allow to fix the slope delta of the diffusion coefficient for some assumed slope of the CR injection spectra (see \cite{Castellina:2005ub} for a detailed discussion about this issue).

In our analysis we will mainly refer to measurements of the secondary/primary flux ratios of several nuclear species (the most relevant are B/C, N/O and sub-Fe/Fe) and the antiproton and ${\bar p}/p$ spectra, performed by several satellite and balloon experiments.

In order to test our code, we firstly study the conventional case of a uniform diffusion coefficient. Afterwards, we will analyze the previously unconsidered case in which $D$ grows exponentially with the distance from the Galactic Plane (GP) and  traces the radial distribution of supernova remnants (SNR).

This paper is structured as follows: in sec.~\ref{sec:diffusion} we give several theoretical and observational motivations for considering the effects of in-homogeneous diffusion on the CR distribution. Moreover, we explain why it is worth studying perpendicular, rather than isotropic diffusion. In sec.~\ref{sec:code} we describe our numerical model and our main assumptions on several variables entering the CR transport equations. In particular, we describe our assumed CR source and target gas distributions, and the nuclear cross section models we exploit. Then, we describe our main results on the physics of CRs, by studying secondary/primary ratios and individual species spectra in sec.~\ref{sec:test}. We will show that the CR spatial distribution can be significantly affected by in-homogeneous diffusion and discuss how this can have relevant consequence for the solution of the CR gradient problem, as we already argued in \cite{Evoli:2007iy}. Although to simulate a detailed
 map of the $\gamma$-ray diffuse emission, suitable to comparison with the one the FERMI observatory
 will soon provide \cite{GLAST}, is beyond the aims of this work, in sec.~\ref{sec:gamma} we will model the longitudinal profile of this emission along the GP showing that under reasonable condition it matches EGRET measurements. Finally, in sec.~\ref{sec:conclusion} we draw our conclusions.

\section{Theoretical and observational motivations for in-homogeneous diffusion models}
\label{sec:diffusion}

Charged particles diffuse in chaotic magnetic fields due to their scattering onto hydro-magnetic fluctuations. The presence of a regular component of the magnetic field, which is the case in the Milky Way, is expected to break isotropy so that spatial diffusion has to be  described in terms of a diffusion tensor $D_{ij}({\bf x})$. According to \cite{Ptuskin:93} this can be conveniently decomposed as
 \begin{equation}
D_{ij}({\bf x}) = \left(D_\bot({\bf x})  - D_\parallel({\bf x}) \right) {\hat B}_i {\hat B}_j  +  D_\parallel({\bf x})  \delta_{ij} + D_A({\bf x})  \epsilon_{ijk} {\hat B}_k \;,
\end{equation}
where ${\hat B}_i$ are the components of the regular magnetic field versor. The symmetric components $D_\parallel$ and  $D_\bot$ are the diffusion coefficients along and perpendicularly to the regular field ${\bf B}_{0}$, while $D_A$ is the antisymmetric (Hall) diffusion coefficient which accounts for the drift due to the interplay of ${\bf B}_{0}$ and CR density gradient. Since $D_A$ is relevant only at very high energies $(E \simgeq 1$ PeV, see e.g.~\cite{Evoli:2007iy,Candia:04}) we will disregard it in the following.

Since diffusion is related to magnetic processes, diffusion coefficients depend on the particle rigidity $\rho = p(E)/Ze$. Moreover, in general $D_\parallel$ and $D_\bot$ depend differently on $\rho$ and on the strength of hydro-magnetic fluctuations. In the quasi-linear theory (QLT)
\begin{equation}
\label{eq:Dpar}
D_\parallel({\bf x}, \rho) \simeq \frac 1 3  v r_L(\rho) ~\mathcal{P}^{-1}(\rho)
\end{equation}
where  $r_L(\rho) = \rho/B_0$ is the Larmor radius and $\displaystyle \mathcal{P}(\rho) \equiv \frac {\int^\infty_{k_{\rm min}(\rho)} \delta B(k)^{2}}{B_0^2}$ is the integral of  the normalized power spectrum of the turbulent hydromagnetic fluctuation over  modes with wavenumber  $k > k_{\rm min}(\rho)   = 2\pi  r_L^{-1}(\rho)$. A power-law behavior $B(k)^2 \propto k^{-\gamma}$ is generally assumed, with $\gamma = 5/3 ~( 3/2)$ for Kolmogorov (Kraichnan) turbulence spectrum implying $D_\parallel  \propto \rho^{1/3\ (1/2)}$.
 In QLT the perpendicular diffusion coefficient is
\begin{equation}
\label{eq:Dperp}
D_\bot({\bf x}, \rho)  \sim D_\parallel({\bf x}, \rho)~\mathcal{P}(\rho)^2 \ll  1\;,
\end{equation}
meaning that diffusion takes place mainly along the regular magnetic field lines.

Although QLT may not be applicable to the conditions presents in the ISM, more realistic computations \cite{Berezinsky:book}
confirmed that expectation founding  $D_\bot \simeq 0.1~D_\parallel$. MonteCarlo simulations of particle propagation in turbulent
fields \cite{Casse:02,Candia:04,DeMarco:07} also found a similar result (although computation time limits allowed to test it only at energies above $100~\TeV$).

What is most relevant here, however, is the different behavior of $D_\parallel$ and $D_\bot$ as a function of the turbulent power. Simulations of propagation in strongly turbulent fields agree with QLT predicting $D_\parallel$ ($D_\bot$) decreasing (increasing) when $\mathcal{P}(\rho)$ increases. It should be noted that if, as it is generally assumed, the CR source distribution can be approximated to be cylindrically symmetric, and the regular field to be purely azimuthal $\bm{B} = (0, 0, B_\phi)$, parallel diffusion plays no  role\footnote{This conclusion is not expected to change significantly if a possible spiral shape and a tiny dipole component of the regular magnetic fields are accounted for (though more complex scenarios have been considered \cite{Breitschwerdt:2002vs}).}.

Clearly, under this approximation and in the absence of an {\it a priori} criterion to fix the normalization and energy dependence of the diffusion coefficients, the substitution of an isotropic diffusion coefficient with $D_\bot$ would produce no physical effects. This conclusion is no more true, however, if the homogeneous diffusion approximation is relaxed and one tries to correlate spatial variations of the relevant diffusion coefficients to those of the hydro-magnetic fluctuation energy density, as $D_\parallel$ and $D_\bot$  have an opposite behavior as functions of $\mathcal{P}(\rho)$.

Observationally, very little is known about the spatial distribution of hydro-magnetic fluctuations in the Galaxy.
There are, however, evidences  both for a longitude \cite{Sun:04} and latitude  dependence of the fluctuation power \cite{Clegg:92,Haverkorn:06}.  Also from a theoretical point of view, $\mathcal{P}(\rho)$  is quite unlikely to be uniform as fluctuations are expected to be correlated, via particle-wave resonant scattering, to CRs which, in turn, are correlated to the non-uniform source distribution.

A radial variation of the diffusion coefficient may have relevant consequences on the CR spatial distribution in the galactic disk. In \cite{Evoli:2007iy} some of us already pointed out that in-homogeneous diffusion may help reconciling the discrepancy between the rather smooth diffuse $\gamma$-ray longitude profile observed by EGRET \cite{Hunter:97}
with the quite steep SNR (the most likely CR sources) radial distribution ({\it CR gradient problem}). That can be understood as a back-reaction effect: a larger CR density nearby sources induces a larger $\mathcal{P}(\rho) $, hence a larger $D_\bot$, which in turn implies a faster CR  diffusion out of those regions (note that the effect would be opposite for $D_\parallel $). In sec.~\ref{sec:gamma} we will discuss this effect
in more details and show its possible  relevance for the $\gamma$-ray angular distribution.

\begin{figure}[tbp]
 \centering
  \includegraphics[scale=0.6]{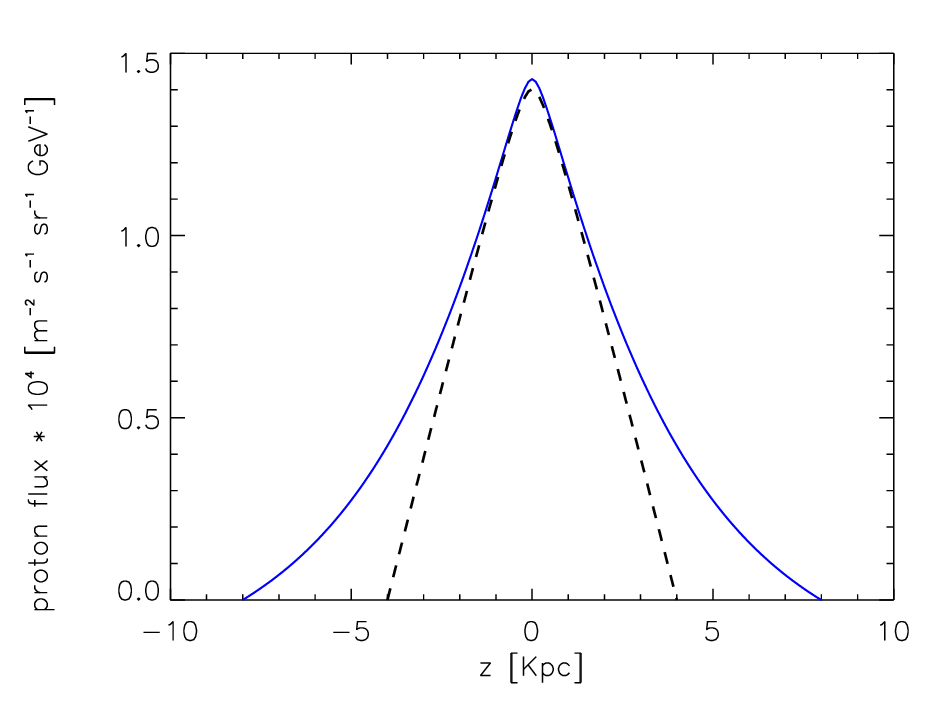}
 \caption{The proton flux vertical profile at 1 GeV  obtained with DRAGON  assuming a uniform diffusion coefficient (black, dashed line) is compared  with that obtained  adopting the exponential profile in equation \ref {eq:exp_D} for $z_t = 4~\kpc$ (blue, continuos line). In both cases  $D$ is normalized so to reproduce the B/C (see section \ref{sec:test}).}
 \label{fig:vert_prof}
\end{figure}

Concerning the vertical profile of the diffusion coefficient, we assume here
\begin{equation}
\label{eq:exp_D}
D(\rho,r,z) = D(\rho,r)~\exp\left\{|z|/z_t \right\}\;.
\end{equation}
This  behaviour, which was also adopted  in \cite{Shibata},  naturally leads to  a more regular  profile of the  CR density  at large $| z |$ (see figure \ref{fig:vert_prof}).
We note, however, that as far as stable secondary nuclei are concerned, replacing an exponentially vertically growing $D_{\bot}$ with a uniform one has almost no effects, as expected because spallation takes place mainly in the thin Galactic disk where the CR density is only marginally affected by the choice between these two options.

%

\section{Description of the model} \label{sec:code}

It is well known since the pioneering work of Ginzburg and Syrovatskii \cite{Ginzburg:64}, that in absence of continuos energy losses, re-acceleration and convection, the diffusive transport of stable nuclei in the ISM is described by the following equation
\begin{eqnarray}
\label{eq:transport}
 \frac{\partial N_i}{\partial t} &+& {\bm\nabla}\cdot\left( \bm{D}\cdot\,{\bm \nabla}N_i \right)
 = \\
&& Q_{i}(E_{k},r,z) - c\,\beta\,n_{\rm gas}(r,z)\,\sigma_{\rm in}(E_{k})N_{i} + \sum_{j>i}c\,\beta\, n_{\rm gas}(r,z)\, \sigma_{ji}N_{j}~.
\nonumber
\end{eqnarray}
where $E_{k} \equiv (E-m_{A})/A$ ($E$ is the total energy of a nucleus with mass $m_{A} \simeq A\times m_{\rm nucleon}$) is  the kinetic energy per nucleon $E_{k}$, which is constant during propagation and is practically conserved in fragmentation reactions.
$\beta$ is the velocity of the nucleus in units of the speed of light $c$.
$\sigma_i$ is the total inelastic cross section onto the ISM gas with density $n_{\rm gas}(r,z) $ and $\sigma_{ij}$ is the production cross-section of a nuclear species $j$ by the fragmentation of the $i$-th one. We start the spallation routine from $A = 64$.
 We disregard ionization and Coulomb energy losses as they are known to be negligible above $\sim1~\GeV$/n. Indeed, we verified with GALPROP that their contribution to the predicted secondary/primary and antiproton spectra is only a few percent above this energy.

We solve equation (\ref{eq:transport}) numerically in the stationary limit $\partial N_{i}/\partial t = 0$ using a Cranck-Nicholson scheme \cite{Press:92,Strong:98,Candia:02,Evoli:2007iy}
with open boundary conditions. This corresponds to free escape of CRs at the outer limit of the Galaxy, defined by $R_{\rm max} = 20~\kpc$ and $z_{\rm max}$.
While $R_{\rm max}$ is fixed, $z_{\rm max}$ is set to $z_{\rm max} = 2\times z_{t}$ (cf.~eq.~\ref{eq:exp_D}) to avoid border effects.

We describe below our assumptions for the terms appearing in eq.~(\ref{eq:transport}).
\begin{description}
\item[Spatial diffusion] The l.h.s.~of eq.~(\ref{eq:transport}) describes spatial diffusion. As we discussed above, we assume cylindrical symmetry and that the regular magnetic field is azimuthally oriented $({\bf B_0} = B_\phi(r,z)\,\hat{\bm{\phi}})$. Under these conditions CR diffusion out of the Galaxy takes place only perpendicularly to ${\bf B_0}$. Therefore in the following it is understood that $D$ represents in fact the perpendicular diffusion coefficient $D_\perp$. The dependence of $D$ on the particle rigidity $\rho$  is (see e.g.~\cite{Ptuskin:97})
\begin{equation}
\label{eq:diff_coeff}
 D(\rho, r,z) = D_0\ f(r)~\beta \left(\frac{\rho}{\rho_0}\right)^\delta\ ~ \exp\left\{|z|/z_t \right\}\;.
 \end{equation}
The function $f(r)$ describes a possible radial dependence of $D$. We define it to be  unity at Sun position $(r = r_\odot )$ so that $D_0$ corresponds to the local value of the diffusion coefficient at the reference rigidity $\rho_0 = 3$ GV.
\item[CR sources] For the source term we assume the general form
\begin{equation}
Q_{i}(E_{k},r,z) =  f_S(r,z)\  q^{i}_{0}\ \left(\frac{\rho}{\rho_0}\right)^{- \alpha_i} \;,
\end{equation}
imposing $Q_{i}(E_{k},r_{\odot},z_{\odot}) = 1$.
While the exact form of $f_{S}(r,z)$ has little influence on the charged secondary species spectra, it is very important in shaping the $\gamma$-ray angular distribution.
We assume $f_S(r,z)$ to trace the SNR distribution as modeled in \cite{Ferriere:01} on the basis of pulsar and progenitor star surveys \cite{Evoli:2007iy}. In the galactic disk such a distribution is similar to that adopted in \cite{Strong:04b}, but shows an excess in the Galactic Bulge due to the contribution of type-Ia SNe, not accounted for in \cite{Strong:04b}. Both distributions are significantly more peaked than those empirically determined \cite{Mattox:96,Strong:98} by matching the $\gamma$-ray longitude profile measured by EGRET \cite{Hunter:97}.

The injection abundances $q^i_0$ are tuned so that the propagated spectra of primary and secondary (or their ratio) species
fit  the observed ones (see below).  Even though our code allows to consider different  power-law indexes $\alpha_i$ for the different nuclear species, in this work we only consider the same $\alpha_{i}\equiv\alpha$ for all species, when not differently stated. For each value of $\delta$ in eq.~(\ref{eq:diff_coeff}) the source spectral slope $\alpha$ is fixed by the requirement that at high energy $E_k \gg 100~\GeV$/n, at which spallation processes are almost irrelevant, the equality $\alpha + \delta = 2.7$ is satisfied\footnote{In this regime, the theoretical expectation for the observed flux $\Phi$ on Earth is $\Phi(E) \approx Q(E)/D(E) \sim E^{-(\alpha+\delta)}$ \cite{Berezinsky:book}.}, in order to match the observed slope. It is understood that DRAGON (similarly to GALPROP) cannot account for the local conditions (e.g.~enhanced turbulence strength and gas density) in the vicinity of the sources, because its limited spatial resolution does not allow to investigate lengths $\simleq 40~\pc$ (while the typical source size is $d \simleq 10~\pc$). These effects, however, are unlikely to produce a significant effect on the CR spectrum,  as the mean fraction of the CR path-length spent in those regions is quite small.
\item[Nuclear cross sections]
%
We included the routines and data tables taken from the public version of GALPROP~\cite{GALPROPweb}. 
In more detail, the material included in our code contains~(see, also,~\cite{SM01}): 
1) The nuclear reaction network, built using the Nuclear Data Sheets;   
2) The isotopic cross section database built using the T16 Los Alamos compilation \cite{t16lib} and the CEM2k and LAQGSM codes~\cite{codes}; 
3) Fits to some particular channels of isotopic production cross section~\cite{MMS01,M03,MM03};
4) phenomenological approximations adapted from~\cite{W-code} and~\cite{newyield}; 
5) inelastic cross section database adapted from~\cite{crosec1,crosec2}.
\item[Target gas] The IS  gas is composed mainly by molecular, atomic and ionized hydrogen (respectively, H$_2$, HI and HII). Although more realistic distributions are known, for $r>2~\kpc$ we adopt the same distributions as in GALPROP, for essentially two reasons. First of all, since CRs propagate for million years in the Galaxy, in the stationary limit they just probe a smoothed, mean gas distribution. Secondly, we can have a more direct comparison with GALPROP results.

However, in the central region of the Galaxy, where GALPROP assumes an interpolated density, we use the the H$_2$ and HI distributions as modeled in \cite{Ferriere:07}. While the flux and composition of charged CR reaching the Earth are not sensitive to the central gas distribution, this choice allows us to better model the $\gamma$-ray emission in the Galactic Centre
(GC) region, as we will discuss in more details in sec.~\ref{sec:gamma}.
Following \cite{Asplund:2004eu} we take the He/H numerical fraction in the ISM to be 0.11. We neglect heavier nuclear species.
\end{description}

\section{Testing DRAGON: the case of a radially uniform diffusion coefficient}
\label{sec:test}

In order to test our code, we ran it under similar conditions to those already considered in the literature. In this section we show the results we obtained assuming that the diffusion coefficient does not depend on the galactocentric radius $r$.
As we mentioned in sec.~\ref{sec:diffusion}, the adoption of an exponential vertical profile for $D$ does not affect significantly the results presented in this section with respect to the case of isotropic and uniform diffusion mostly considered in literature. Indeed, passing from a spatially uniform $D$ to the profile described by eq.~(\ref{eq:diff_coeff}) only amounts to a small re-scaling of $D_0$.

In the following, every label indicating a nucleus refers in fact to the sum of all its isotopes, unless otherwise stated.

\subsection{The B/C ratio}\label{sec:BvsC}

The Boron to Carbon ratio (B/C) is one of the most useful tracers of CR propagation in the Galaxy. In fact, since Boron is entirely secondary, its observed abundance strongly depends on the residence time of primary CRs in the Galaxy. Moreover, measurements of Boron and Carbon fluxes are better than those of other secondary/primary ratios, and the B production cross sections from its main primaries ($^{12}$C and $^{16}$O) are known better than for other secondary nuclides.

\subsubsection{Fixing free parameters}

Once the spatial distributions of the CR sources and the ISM gas have been chosen, the  main parameters determining the B/C in a PD model are the C/O and N/O injection ratios 
 and the quantities $\delta$, $D_0$ and  $z_t$ in eq.~(\ref{eq:diff_coeff}). As it was already shown in several papers,
 secondary/primary ratios for stable species depend on the ratio $D_{0}/z_{t}$  (which will be always expressed in units of
 $10^{28}~\cm^2~\s^{-1}~\kpc^{-1}$ throughout this paper) rather than on the two parameters separately.

While primary/primary ratios are usually disregarded in the literature, as they do not give direct relevant information on CR propagation, we use them to fix the C/O and N/O\footnote{Note that N = $^{14}$N +  $^{15}$N is a combination of primary
and secondary nuclides} injection ratios, while we fix the abundances of primaries heavier than oxygen by requiring that they match the observed abundances in CRs at $E \sim 1-10~\GeV$/n.

To this aim, we define two different $\chi^{2}$. We compute the former (which we label $\chi^{2}_{\rm \{C/O, N/O\}}$) by comparing our predictions for the C/O and N/O modulated ratios to experimental data over the energy range of our interest. The latter (which we label $\chi^{2}_{\{D_{0}/z_{t},\,\delta\}}$) is computed comparing our predicted modulated B/C ratio to the observed one. Solar modulation is taken into account here in the ``force-field" approximation \cite{Gleeson&Axford} using a modulation potential of magnitude $\Phi = 500~\MV$.
In order to study potential energy dependent effects we consider two different minimum kinetic energies per nucleon $E_{k}^{\rm min}$ for comparison to data: 1 GeV/n and 2 GeV/n. The low statistical significance of the data set above this energy prevents us from going further up the energy scale.

For each pair of values ($D_0/z_t$, $\delta$), we determine the $\chi^{2}_{\rm \{C/O, N/O\}}$ distribution in the space (C/O, N/O) scanning over a wide range of C/O and N/O injection ratios. For the set of parameters that minimizes $\chi^{2}_{\rm \{C/O, N/O\}}$, we compute $\chi^{2}_{\{D_{0}/z_{t},\,\delta\}}$ and we repeat this procedure for several values of ($D_{0}/z_{t},\delta$).  Finally, we analyze the distribution of $\chi^{2}_{\{D_{0}/z_{t},\,\delta\}}$ to obtain our best fit values for $(D_{0}/z_{t},\,\delta)$ with the appropriate confidence regions.

Thus, this strategy allows us to fix best values of the C/O and N/O injection ratios and to consistently determine the best propagation parameters that will be used as our best model for the analysis of antiproton and $\gamma$-ray fluxes.

We notice here that this procedure, which corresponds essentially to split the whole 4$-D$ parameter space into two separate ones, is physically motivated by the weak dependence of primary/primary ratios on ($D_{0}/z_{t}, \delta$).

\subsubsection{Experimental data}

So far the best B/C measurements above $1~ \GeV$/n have been provided by the HEAO-3 \cite{HEAO-3} and CRN \cite{CRN} experiments in the range $1 < E_k < 30~\GeV$/n and $ 70~\GeV\mathrm{/n}< E_k \simleq 1.1~\TeV$/n. New data should be released soon by the CREAM \cite{CREAM} and TRACER \cite{TRACER} experiments significantly improving the available statistics at high energy. Here we use only HEAO-3, CRN data.
For consistency, we take also C/O and N/O data from the same experiments.

\subsubsection{Results}
\label{sec:BC}

\begin{figure}[tbp]
 \centering
 \subfigure[]{
   \label{fig:chi2}
   \includegraphics[scale=0.45]{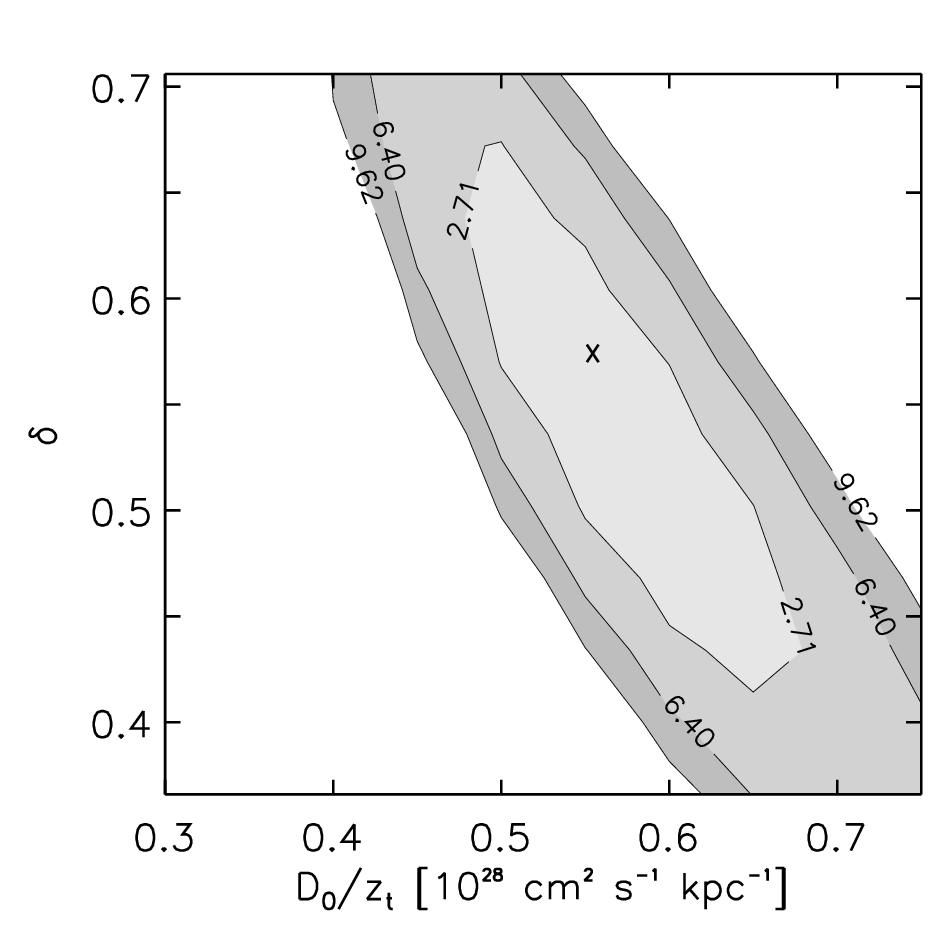}
 }
 \subfigure[]{
   \label{fig:chi22GeV}
   \includegraphics[scale=0.45]{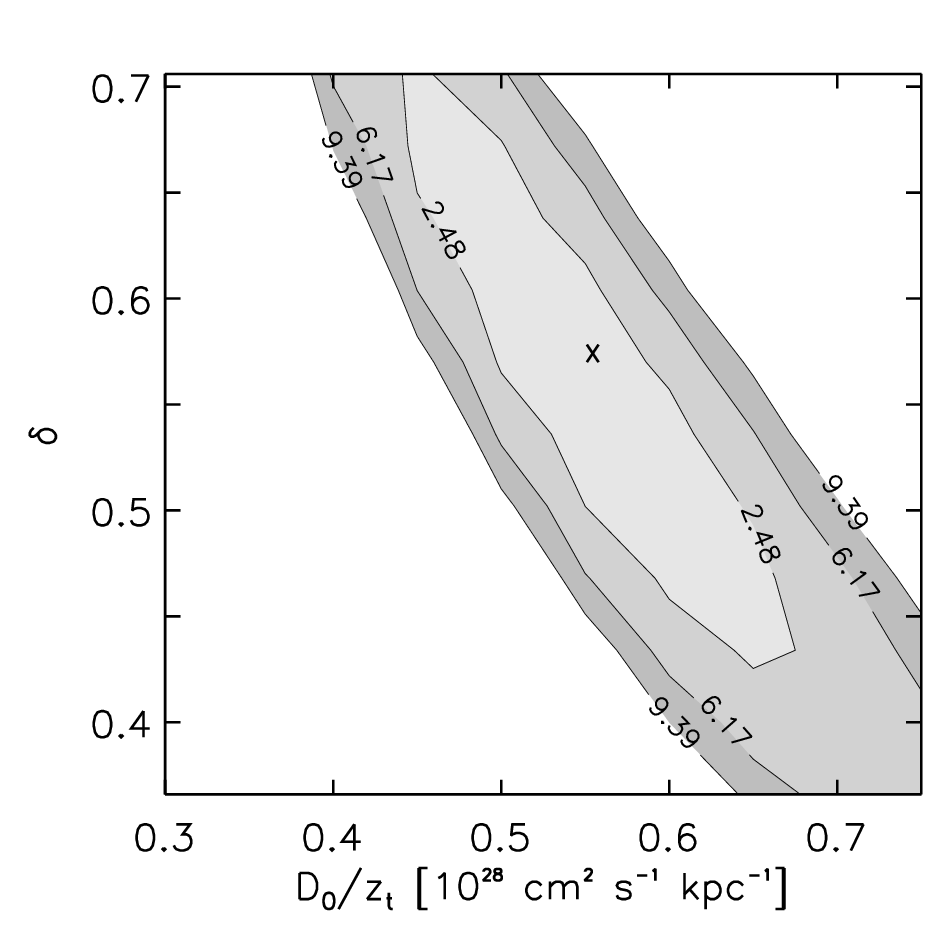}
 }
 \caption{The distribution of $\chi^{2}_{\{D_{0}/z_{t},\,\delta\}}$ is shown for the case $E_{k}^{\rm min} = 1$ GeV/n (left) and 2 GeV/n (right). Contours limit 1, 2 and 3 $\sigma$ confidence regions. }
\end{figure}

We show in figures \ref{fig:chi2} and \ref{fig:chi22GeV} the main results of our procedure. In fig.~\ref{fig:chi2} the distribution of $\chi^{2}_{\{D_{0}/z_{t},\,\delta\}}$ for $E_{k}^{\rm min} = 1~\GeV$/n is shown, together with confidence regions at 68\%, 95\% and 99\% Confidence Level (CL).
Our best-fit values for ($D_{0}/z_{t}, \delta, \mathrm{C/O, N/O}$) if $E_{k}^{\rm min} = 1$ GeV/n are (0.55, 0.57, 0.79, 0.044).
The projection of this point in the ($D_{0}/z_{t}, \delta$) plane is highlighted by the cross in figure \ref{fig:chi2}. Remarkably, the best-fit value for $\delta$ favors a Kraichnan turbulence spectrum, rather than a Kolmogorov one. Changing the minimum energy $E_{k}^{\rm min}$ from 1 GeV/n to 2 GeV/n indeed produces no relevant effect. In particular, the best-fit values for $D_{0}/z_{t} $ and $\delta $ are not moved (see fig.~\ref{fig:chi22GeV}). It is interesting to notice that the particular value of $\delta = 0.57$ we obtain is consistent with findings of other authors (see \cite{Strong:2007nh} and references therein).
The best-fit C/O and N/O injection ratios (0.79,0.044)   should be compared with the solar system ones \cite{Asplund:2004eu}
0.76 and 0.11 respectively.

\begin{figure}[tbp]
 \centering
 \includegraphics[scale=0.45]{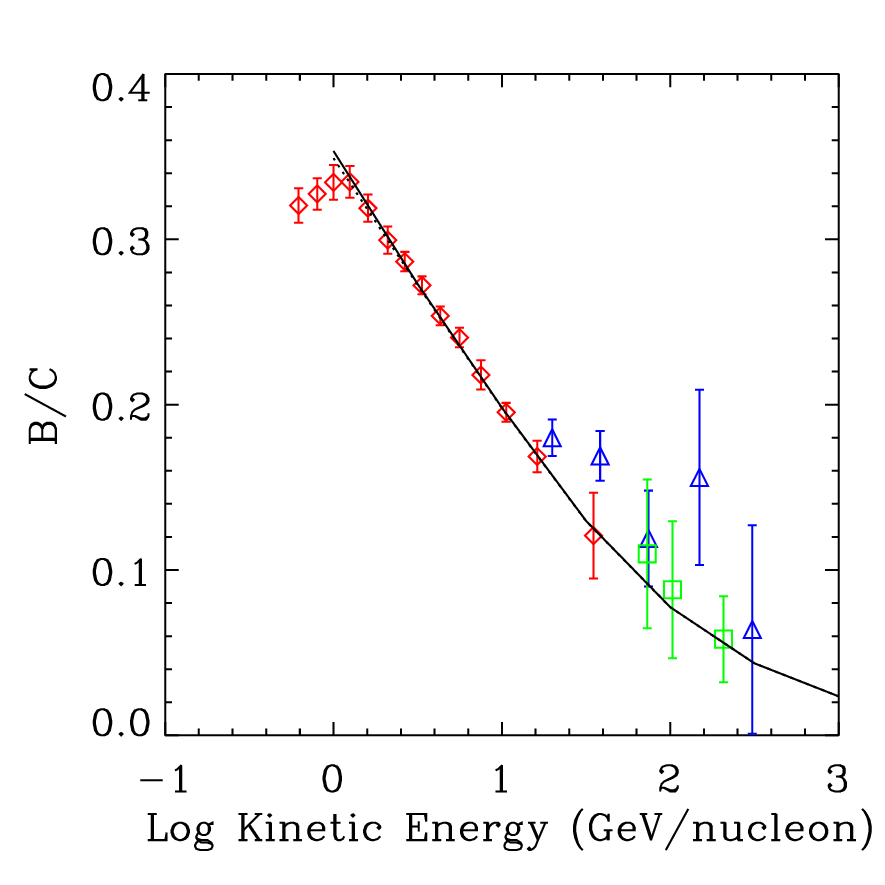}
 \includegraphics[scale=0.45]{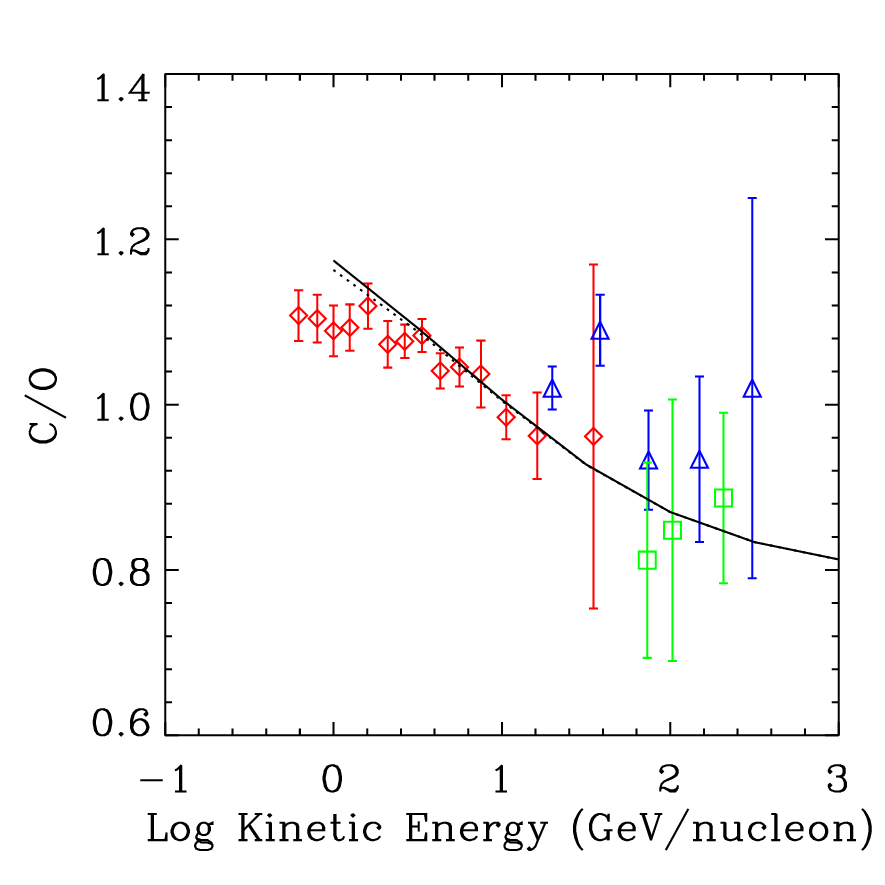}
 \includegraphics[scale=0.45]{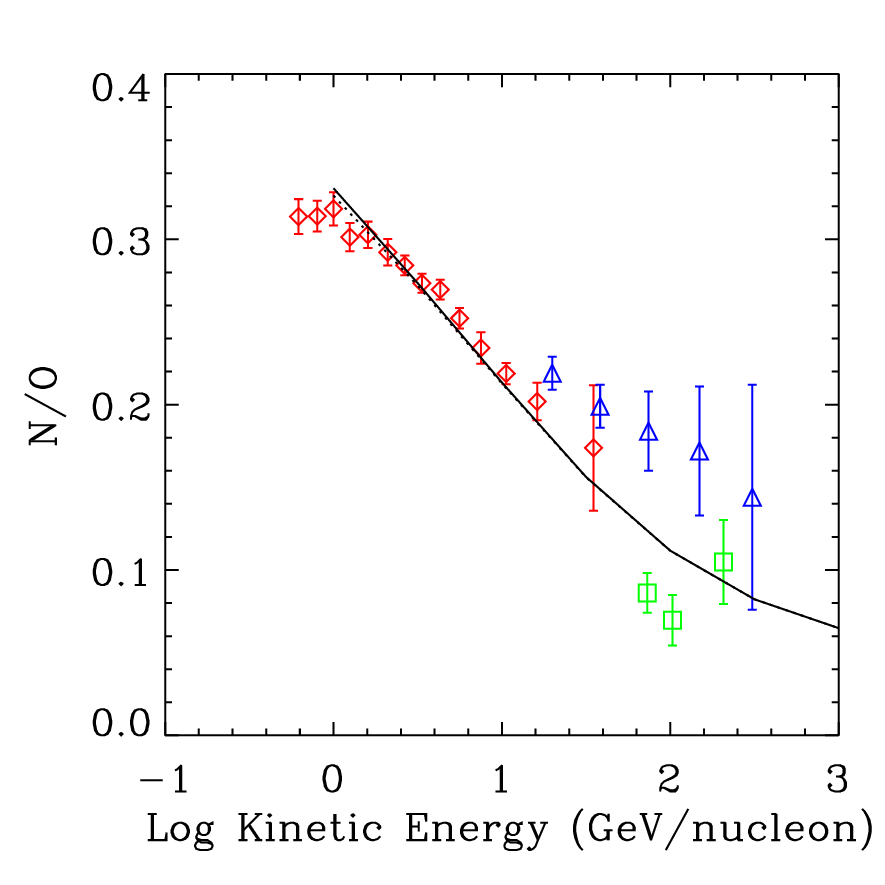}
 \caption{In these panels we show our best fit for the B/C, C/O and  N/O compared with
 HEAO-3 \cite{HEAO-3} (red diamonds), CRN \cite{CRN} (green, triangle) and ATIC-II \cite{ATIC} (blue) experimental data
 (though the latter are not used in our statistical analysis).
 Continuos curves:  local interstellar  (LIS);   dashed lines: top of atmosphere (TOA)  ($\Phi = 500~\MV)$.
 }
 \label{fig:best_fit}
\end{figure}

We do not include the sub-Fe/Fe (sub-Fe = V + Ti + Sc) ratio in our statistical analysis because of the large uncertainties on the knowledge of the spallation cross sections for heavy elements. However,
we found that we consistently obtain a reasonable match of experimental data also for this observable. In order to improve
the fit to this ratio a careful fine tuning of nuclear cross section parametrizations seems to be needed.
In fig.~\ref{fig:best_fit} we show the B/C, C/O and N/O ratios as obtained with our best-fit model, and assuming $z_t = 4~\kpc$.

A comment is in order here: the particular observables we are considering are not sensitive to $D_{0}$ and $z_{t}$ independently. A possible way to estimate $z_{t}$ is offered by unstable/stable ratios (e.g.~$^{10}$Be/$^{9}$Be), which are known to probe the vertical height of the Galaxy \cite{Berezinsky:book}. Unfortunately, the best experimental data for this particular ratio have been obtained at energies $\simleq 100$ MeV/n, while only 2 experimental points with large errors are available at 1 GeV/n \cite{ISOMAX}. Since our code does not allow us to have reliable predictions down to few hundreds MeV/n, because we do not account for energy losses, it is impossible for us to draw any firm conclusion about our best value for $z_{t}$. However, by considering only the $^{10}$Be/$^{9}$Be points around 1 GeV/n we infer that $z_{t}$ should lie between 3 and 5 kpc, in agreement with previous findings \cite{Moskalenko:2001qm}.
%
%

\subsection{Antiprotons}\label{sec:antiprotons}

Most  antiprotons reaching the Earth are expected to be a product of CR hadronic collisions with the IS gas. Their measured spectra provide, therefore, valuable information on CR propagation which are complementary to that coming from secondary nuclei (see e.g. \cite{Bergstrom:1999jc,Donato:01,Moskalenko:2001ya}).

The main processes responsible  for ${\bar p}$ production are $p\; p_{\rm gas}$, $p \;{\rm He}_{\rm gas}$, ${\rm He}\; p_{\rm gas}$ and ${\rm He}\; {\rm He}_{\rm gas}$, plus a negligible contribution from other nuclei. Similarly to \cite{Donato:01,Moskalenko:2001ya} we use the ${\bar p}$ production cross-section calculated using the parametrization given in Tan \& Ng \cite{Tan:1982nc}. We account for the contribution of heavier nuclei in the CRs and the ISM by using the effective correction function determined by Simon {\it et al.}~\cite{Simon:98} with the MonteCarlo model DTUNUC. Inelastic scattering, annihilation and tertiary ${\bar p}$ (antiprotons  which have been inelastically scattered) are treated as in \cite{Moskalenko:2001ya}. For the local interstellar spectrum (LIS) of primary nuclei we adopt $\Phi_p = 1.6 \times 10^4\ (E_k/1~\GeV)^{-2.73}~
(\m^2~\s~ \sr~ \GeV)^{-1}$ as measured by BESS during the 1998 flight \cite{Sanuki:2000wh} by accounting for a solar modulation potential $\Phi = 550~\MV$ in the ``force-free" approximation \cite{Gleeson&Axford}.

We use DRAGON to simulate the primary proton distribution in the Galaxy and the LIS  of secondary antiprotons. Normalization is imposed by requiring that the simulated proton LIS coincides with $\Phi_p(r_\odot, 0)_{\rm obs}$\footnote{Only the absolute ${\bar p}$ flux is  dependent on such normalization.}.
\begin{figure}[tbp]
 \centering
 \subfigure[]
 {
 \label{fig:ap_ratio}
 \includegraphics[scale=0.45]{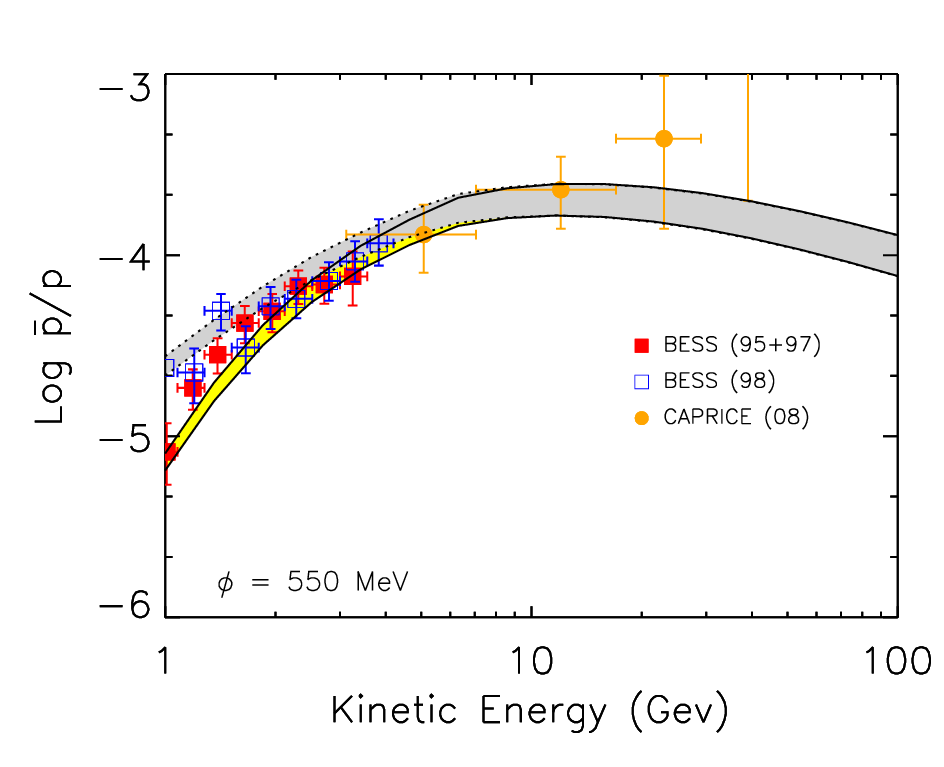}
  }
  \subfigure[]
  {
   \label{fig:ap_spectrum}
  \includegraphics[scale=0.45]{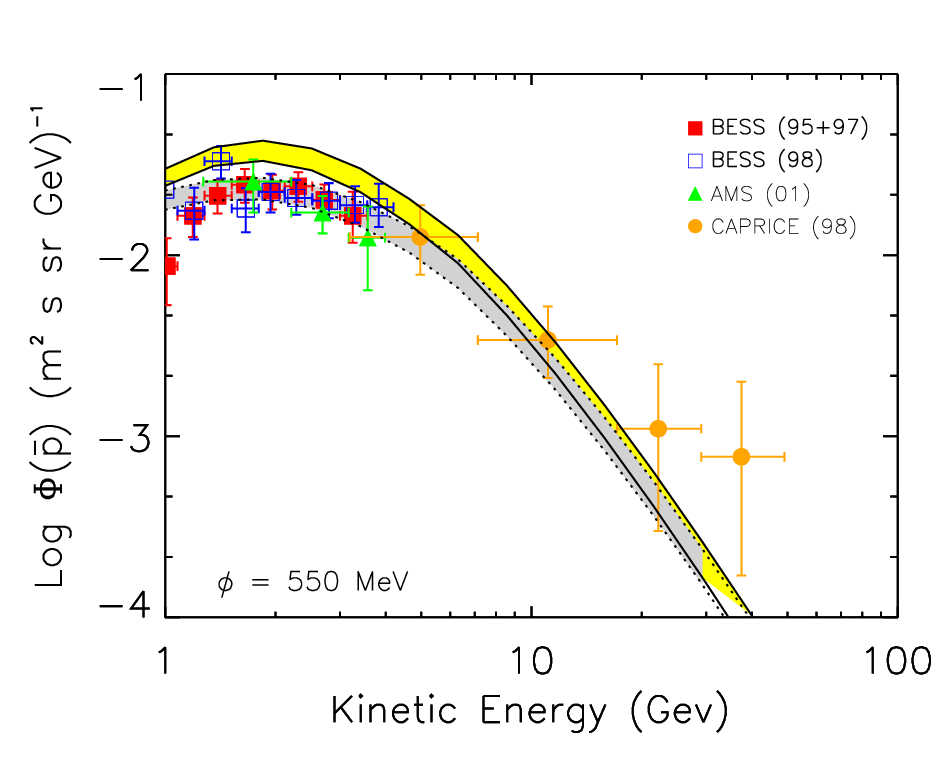}
  }
 \caption{The ${\bar p}/p$ ratio (left) and antiproton absolute spectrum (right) are compared with  BESS 95+97 \cite{Orito:1999re}, BESS98 \cite{Maeno:2000qx}, CAPRICE  \cite{Boezio:2001ac}, and AMS \cite{AMS}  experimental data.
 The shadowed  regions correspond to models matching the B/C data within $1 \sigma$.
LIS: light yellow band, between continuos lines;  modulated  $(\Phi = 550~\MV)$: grey band, between dashed lines.
 }
\end{figure}
In figures \ref{fig:ap_spectrum} and \ref{fig:ap_ratio} we compare our results with the experimental data released by BESS for the periods 1995-97 \cite{Orito:1999re} and 1998 \cite{Maeno:2000qx} in the energy interval $1-4~\GeV$, and by CAPRICE (1998) \cite{Boezio:2001ac} in the range $3-49~\GeV$. All these data refer to a period of low solar activity (the minimum was in 1997) and same positive phase of the solar cycle, with a mean value of the modulation potential for the period 1995-1998 of $\Phi = 550~\MV$ \cite{Maeno:2000qx}. Hence we will also use $\Phi = 550~\MV$ to obtain $\bar{p}$ modulated spectra.

The shaded regions correspond to the uncertainty on the antiproton flux due to the uncertainty on the propagation parameters, and are constructed using values of ($D_{0}/z_{t}, \delta$) within the $1\sigma$ region of fig.~\ref{fig:chi2}. The figures shown in this section are derived using  $z_t = 4~\kpc$, but we checked that, similarly to the B/C case, different choices of this parameter do not affect the antiproton spectrum provided that $D_{0}/z_{t}$ is kept constant.

In fig.~\ref{fig:ap_ratio} we also show the LIS and TOA energy behavior of the ${\bar p}/p$ ratio obtained with the parameters corresponding to minimum  of the $\chi^2$ distribution shown in fig.~\ref{fig:chi2}. It is evident that the models which fit the B/C data within $1\sigma$ are also compatible with the antiproton measurements.
A statistically poor excess of the predictions of our  best-fit model respect to the BESS data, which was also found in \cite{Moskalenko:2001ya}, is probably not significant due to  the large systematic uncertainties. At higher energies, we have a small  tension between our predictions and the highest energy CAPRICE data. A better agreement may be found if  preliminary PAMELA \cite{PAMELA} results \cite{PAMELA:web} will be confirmed.

\section{Radial dependent diffusion and the $\gamma$-ray longitude distribution}
\label{sec:gamma}

In this section we model the secondary $\gamma$-ray emission originated, via $\pi^{0}$ decay, by the interaction of the hadronic component of CRs with the IS gas. Along the GP, where the gas column density is higher, this process is expected to give the dominant contribution to the total diffuse emission above the GeV.
At the energies of our interest a simple scaling model for the differential production cross section can still reliably be used. In this regime the energy spectrum of secondary $\gamma$'s is a power law with the same slope as the primary nuclei
(only protons and He nuclei give a significant contribution).

The main gas (target) components are the molecular (H$_2$) and atomic (HI) hydrogen, and He atoms. The contribution of ionized hydrogen is almost irrelevant in the GP. For $r > 2~\kpc$, we adopt the same HI spatial distribution as \cite{Strong:98}. For the ${\rm H}_2$ we assume
\begin{equation}
n_{{\rm H}_2}(r,z) = \epsilon_0(r) ~X_{\rm CO}(r)~ \exp\left\{- \ln 2 (z - z_0)^2/z_h(r) \right\},
\end{equation}
where $\epsilon(r)$ is the CO (a widely used ${\rm H}_2$ tracer) volume emissivity, $z_0(r)$ and $z_h(r)$ are the midplane displacement and scale heights respectively, and $X_{\rm CO}(r)$ is the CO - ${\rm H}_2$ conversion factor. All these quantities, with the exception of $X_{\rm CO}(r)$, are the same as in \cite{Bronfman:88,Strong:98,Strong:04} for $r > 2~\kpc$, while for smaller radii we adopt the Ferriere {\it et al.}~model \cite{Ferriere:07}. The adoption of Ferriere's model for the molecular and atomic hydrogen for $r < 2~\kpc$ allows us to avoid the interpolation of the $\gamma$-ray flux profile in the GC region and to reproduce naturally the peaked emission observed by EGRET toward the GC as we already pointed out
in \cite{Evoli:2007zz}. For the $11\% $ He fraction we adopt the same spatial distribution as for the HI.

\subsection{The CR gradient problem}

To simulate a detailed map of the $\gamma$-ray diffuse emission of the Galaxy (even only for the hadronic component) is beyond the aims of this work. Rather, the main issue we want to address here is the so called {\em CR gradient problem}. This originates from the well known discrepancy between the theoretical flux profile obtained by assuming SNRs to be the sources of galactic CRs and that inferred from EGRET $\gamma$-ray diffuse observations \cite{Hunter:97}. Under mild assumptions on the distribution of the galactic gas, it was found \cite{Mattox:96,Strong:98} that the inferred CR radial profile should be much flatter than the theoretically expected one.

\subsubsection{A former proposed solution and a test for DRAGON}

A possible way out was suggested in \cite{Strong:04b} in terms of a radially variable $X_{\rm CO}$. While in \cite{Strong:98,Strong:04} this quantity was assumed to be uniform ($X_{\rm CO} = 1.8 \times 10^{20}~\cm^{-2}/({\rm K~km} \s^{-1})$)\footnote{For clarity, in the following we will drop units in quoting values of $X_{\rm CO}$. They are always understood to be $\cm^{-2}/({\rm K~km}\, \s^{-1})$.}, in \cite{Strong:04b} it was taken to increase gradually by more than one order of magnitude from $4 \times 10^{19}$ at $r = 2~\kpc$ to $1 \times 10^{21}$ at $r > 10 ~\kpc$. However, while the growth of this parameter with $r$ is suggested both by theoretical arguments and observations of external galaxies, its actual behavior is rather uncertain so that in \cite{Strong:04b} it had to be tuned into 5 steps to match EGRET observations.

To test our code against possible failures in reproducing the $\gamma$-ray longitude profile, we try to reproduce the results of \cite{Strong:04b}. We adopt the same $X_{\rm CO}(r)$ which was used in \cite{Strong:04b} and a CR model giving the best-fit of the
B/C in the case of a radially uniform diffusion coefficient. We use $\delta = 0.57$ (see sec.~\ref{sec:BC}) but our results do not change appreciably by using any value in the interval $0.45 - 0.65$. In fig.~\ref{fig:gamma_const} we compare our results with EGRET measurements along the GP  for $4 < E_\gamma < 10~\GeV$ \cite{egret}.

We reasonably reproduce both the normalization and the main features of the observed longitude profile. Smaller structures may only be reproduced using a detailed 3$-D$ model of gas distribution which we are planning to do in a forthcoming paper. For comparison, in the same figure  we also show the emission profile which we would obtain using a constant $X_{\rm CO}(r) = 1.8 \times 10^{20}$ for $r > 2~\kpc$.

\subsection{An alternative solution of the CR gradient problem}

As an alternative possibility we explore the case in which the diffusion coefficient  traces the radial dependence of the SNR distribution as we motivated in sec.~\ref{sec:diffusion}. According to the arguments explained in the same section we expect
the CR radial profile to be smoothed with respect to the one obtained in the case of constant diffusion coefficient. Hence, we expect to be able to fit EGRET longitude profile without fine tuning the parameter $X_{\rm CO}$. Indeed, this is what we find.

We assume a constant $X_{\rm CO} = 1.8 \times 10^{20}$ for $r > 2~\kpc$, while in the bulge ($r < 2$ kpc), where physical conditions are much different from the outer disk, we take  $X_{\rm CO} = 0.5 \times 10^{20}$ \cite{Ferriere:07}. For the diffusion coefficient, we assume that the function $f(r)$, as defined in eq.~(\ref{eq:diff_coeff}), is
\begin{equation}
\label{eq:tau_def}
f(r)  = f_S(r,0)^{\tau}\;.
\end{equation}
The function $f_S(r,0)$ describing the radial distribution of the Galactic SNRs is taken by \cite{Ferriere:01} and is the same as in \cite{Evoli:2007iy}.

A self-consistent model should be developed in order to rigourously determine $f(r)$ and the primary CR density $N_i(r)$ accounting for the mutual influence of CR and  hydromagnetic fluctuations.  Although such an analysis is beyond the aims of this work, a simple estimate can be derived under reasonable physical assumptions. In particular, we assume that, as observed locally and expected by theoretical reasons, energy equipartition holds between CR and magnetic field fluctuations. Under this hypothesis, and restricting our analysis to the galactic disc, we have that
the fluctuation strength $\mathcal{P}$ is proportional to the CR density $N(r,0)$, and, since equations (\ref{eq:Dpar}) and (\ref{eq:Dperp}) imply $D_\perp(r) \propto \mathcal P$, we obtain $D_{\perp}(r) \propto N(r,0)$ (we further simplify the problem by assuming that $D_{\perp}$ is constant in $z$, since, as we also pointed out in the previous section, our observables do not depend strongly on the vertical behaviour of $D_{\perp}$).
Assuming now that the source density is of the form $f_S(r,z) \simeq g(z)F(r)$, it is straightforward to show that, in the limit in which radial diffusion can be neglected with respect to the vertical one (or, in other words, that the diffusion halo radius is much larger than is vertical thickness), the stationary limit of the spatial diffusion equation  takes the form $\displaystyle N(r,0)\cdot\frac{\partial^2 N(r,z)}{\partial z^2} = K g(z)F(r)$, where $K$ is a constant. This equation can be easily solved by assuming $N(r,z) = N(r,0)H(z)$, from which we obtain
\begin{equation}
D_{\perp} \propto N(r,0) = F(r)^{0.5}\;,
\end{equation}
implying $\tau = 0.5$.

\begin{figure}[tbp]
\centering
\includegraphics[scale = 0.5]{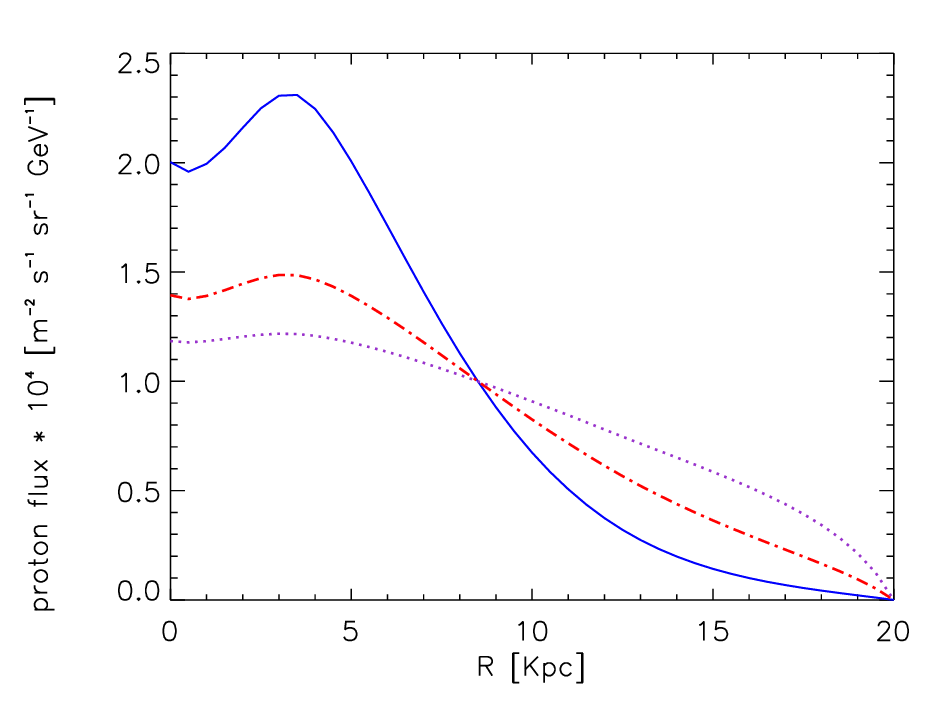}
\caption{Proton differential flux at $E = 1~\GeV$ for three different choices of the parameter $\tau$ setting the radial dependence of the diffusion coefficient on the SNR distribution (see eq.~(\ref{eq:tau_def})). $\tau = 0$ (radially uniform $D$), blue, continuous curve; $\tau = 0.5$, red, dash-dotted;  $\tau = 0.75$, violet, dotted. In all cases $z_t = 4~\kpc$ and the $D$ normalization giving the best fit to B/C data is chosen.}
\label{fig:rad_prof}
\end{figure}

Of course, this rough estimate cannot replace a more detailed self-consistent numerical computation. It is intriguing, however, that, besides its pedagogical utility, this simple derivation predicts a CR radial distribution which is quite close to what is needed to solve the gradient problem.
In figure \ref{fig:rad_prof} we show the radial profile of the proton differential flux at 1 GeV for three different values of the parameter $\tau$: 0,75,  0.5  and 0  (the latter corresponds to a radially uniform $D$).  It is clear that for $\tau \simgeq 0.5$ a  significant flattening  of the CR spatial distribution with respect to that of sources has to be expected.

In order to verify if such distributions are compatible with CR measurments, we run DRAGON under the same conditions discussed in sec.~\ref{sec:test} but adopting  a radially dependent $D$ as specified in (\ref{eq:tau_def}). We find that, as long as $\tau < 1$, we are still able to obtain a good fit of secondary/primary ratios and antiproton data. In particular, we find that for $\tau \simeq  0.5$ this happens with the same propagation parameters which allow to match observations in the case $\tau = 0$, in spite of the fact that those values  correspond to quite different CR radial distributions.

\begin{figure}[tbp]
  \centering
 \subfigure[]
 {
 \label{fig:gamma_const}
 \includegraphics[scale=0.45]{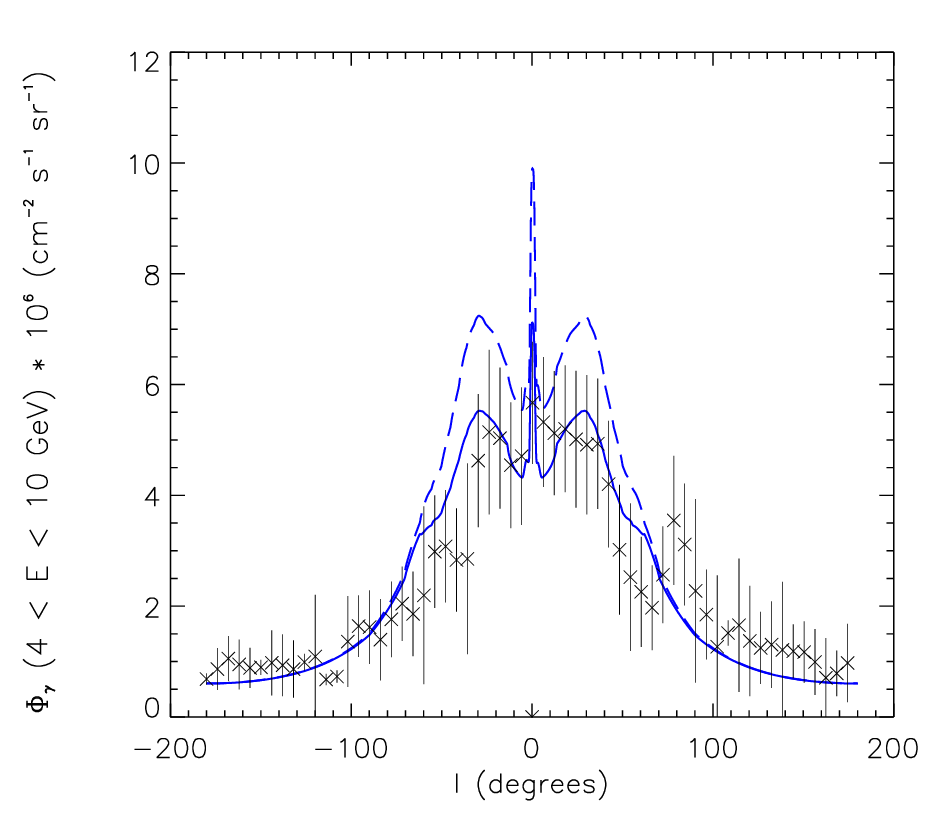}
  }
  \subfigure[]
  {
   \label{fig:gamma_rad}
    \includegraphics[scale=0.45]{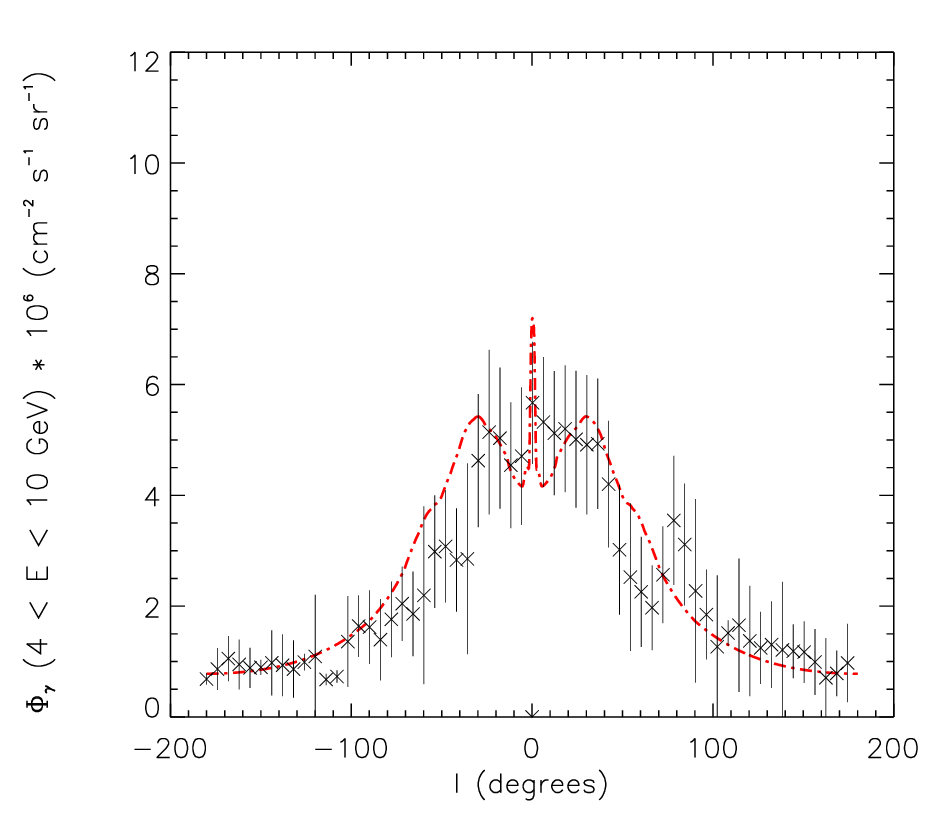}
  }
 \caption{Our predictions for the longitudinal profiles of the $\gamma$-ray hadronic emission integrated for $|b| < 1^\circ$ are compared with EGRET measurements \cite{Hunter:97}.   Left panel: radially uniform $D$ with $X_{\rm CO} =  1.8 \times 10^{20}~\cm^{-2}/({\rm K~km} \s^{-1})$ for $r > 2~\kpc$ (long dashed curve), and $X_{\rm CO}$ as in \cite{Strong:04b} for $r > 2~\kpc$ (continuous curve). Right panel: $D(r)$ tracing the SNR distribution with $\tau = 0.5$. In both cases $z_t = 4~\kpc$ and $D$ normalization is chosen to best-fit the B/C data.}
\end{figure}

The degeneracy of the different CR density profiles with respect to CR measurements could be removed by means of $\gamma$-ray observations if combined with independent astrophysical measurements of the $X_{\rm CO}$ radial dependence.
In fig.~\ref{fig:gamma_rad} we show the simulated $\gamma$-ray longitude profile as obtained using $\tau = 0.5$  and   $X_{\rm CO}$ taken to be uniform for $r > 2~\kpc$.  Clearly this is quite smoother than the profile obtained under the same conditions but a radially uniform $D$ ($\tau = 0$). A slightly better fit of data can be obtained using $\tau = 0.75$, probably indicating that also radial diffusion should be taken into account respect to our semplified  analytical model .
The comparison of figures \ref{fig:gamma_const} and \ref{fig:gamma_rad} displays the degeneracy between the radial dependence of $X_{\rm CO}$ and that of the diffusion coefficient.  It is intriguing, however, that the EGRET profile may be explained in terms of in-homogeneous diffusion with no need to invoke a  tuning of  the $X_{\rm CO}$ parameter. While the actual $\gamma$-ray longitude profile is likely to be determined by a combination of the radial dependence of both those quantities, it should be clear from our results that the effect pointed out in this work should be taken into account when interpreting observations of $\gamma$-ray diffusion emission of the Milky Way and of external galaxies.

\section{Conclusions}
\label{sec:conclusion}

We modeled CR transport in the Galaxy assuming a plain diffusion model. We study two main cases: in the first one the diffusion coefficient $D$ is assumed to be uniform along the Galactic Plane, while in the second case we consider, for the fist time, a $D(r)$ which traces the radial profile of SNRs (which we assume to be the CR sources).

The fact that we can consistently reproduce the observed antiproton spectra and the main secondary/primary nuclear ratios for  $E \simgeq 1~\GeV$/n makes us quite confident of the validity of our approach. By using only the B/C, C/O and N/O data we found that the preferred range ($1\sigma$) of  values of the slope of diffusion coefficient is 0.43-0.65. The best fit value is  $\delta \simeq 0.57$. This is in agreement with findings of other authors.
A Kolmogorov spectrum is disfavored and re-acceleration seems to be unnecessary to interpret data above $1-2~\GeV$/n. Forthcoming experiments like CREAM \cite{CREAM} and TRACER \cite{TRACER}  for what concerns nuclei and PAMELA \cite{PAMELA} and AMS \cite{AMS} for antiprotons may soon allow to strengthen this conclusion by improving both statistics and quality of data.

While in both cases (uniform and radially dependent $D$) we obtain substantially the same successful predictions for what concerns nuclei and antiprotons reaching the Earth, the corresponding CR primary spatial distributions in the Galaxy can be considerably different. This may have a number of interesting effects, including a possible role in the solution of the problem which  plain diffusion models face predicting a too high CR anisotropy above 100 TeV.

We focused here on the effects on the expected secondary $\gamma$-ray diffuse emission. We showed that the longitude distribution of that emission can be significantly affected by in-homogeneous diffusion. In \cite{Evoli:2007iy} we already noticed that the effect goes in the right direction to provide a viable solution of the CR gradient problem.  Here we provide further theoretical arguments in favour of a radially dependent diffusion coefficient and succeed reproducing EGRET observations for $4 < E < 10~\GeV$ and $|b| < 1^\circ$ for a choice of the relevant diffusion parameters which allow to match the B/C and the antiproton spectrum.
 The extension of our predictions to larger latitudes would require to implement in DRAGON electron propagation (and losses) and more detailed gas and radiation distributions which we plan to do in a forthcoming paper.

\section*{Acknowledgements}
We are indebted with P.~Ullio for many valuable suggestions and discussions.
We also thanks P.~Blasi, F.~Donato, J.~Kirk, I.~Moskalenko, S. Shore, A.~Strong and the anonymous referee  for several useful comments. D. Grasso has been supported by ``Agenzia Spaziale Italiana"  (ASI)  in the framework of the Fermi Gamma-Ray Space Telescope project and, partially,   by the EU FP6 Marie Curie Research \& Training Network ``UniverseNet"
(MRTN-CT-2006-035863).

\end{document}